\documentclass{aa}
\input psfig.sty

\def\Msun{$M_\odot$}
\def\simgt{\lower.5ex\hbox{$\; \buildrel > \over \sim \;$}}
\def\simlt{\lower.5ex\hbox{$\; \buildrel < \over \sim \;$}}

\begin{document}

  \thesaurus{06 
	     ( 08.16.4;  
	       08.05.3;  
	       08.13.2)} 
   \title{Full spectrum of turbulence convective mixing: II.
	  Lithium production in AGB stars.}

   \author{I. Mazzitelli \inst{1} F. D'Antona \inst{2} \and
	   P. Ventura \inst{2}}

\offprints{P. Ventura}

   \institute{Osservatorio Astronomico di Roma,
	      Monte Porzio Catone, I-00040 (Rome)\\
	      email: paolo@coma.mporzio.astro.it}

 \date{Received 23 february 1999; accepted ???????}

   \titlerunning{Lithium production in AGB stars}

   \maketitle

   \begin{abstract}

We present results from new, detailed computations of lithium production by
hot bottom burning (HBB) in asymptotic giant branch (AGB) stars of
intermediate mass ($3.5 \leq \rm M \leq 6 M_\odot$). The dependence of
lithium production on stellar mass, metallicity, mass loss rate, convection
and overshooting are discussed. In particular, nuclear burning, turbulent
mixing and convective overshooting (if any) are self--consistently coupled by
a diffusive {\bf algorithm}, and the Full Spectrum of Turbulence (FST) model of
convection is adopted, with test comparisons to Mixing Length Theory (MLT)
stellar models. All the evolutions are followed from pre--main sequence down
to late AGB, when stars do not appear any longer lithium rich. A ``reference
mass" of 6 \Msun\ has been chosen since, although relatively close to the
upper limit for which degenerate $^{12}C$ ignition occurs, all the studied
mechanisms show up more clearly.

HBB is always found above $\sim \log L/L_\odot = 4.4$, but the range of
(initial) masses reaching HBB is largely dependent on convection model,
overshooting and metallicity. For solar chemistry, masses $\geq 4$\Msun\
evolve through HBB in the FST case and including core overshooting whereas,
with solarly tuned MLT models and no overshooting, only masses $\geq
6$\Msun\
can reach HBB. These constraints can give feedbacks about the more correct
convection model and/or the extent of overshooting, thanks to the signatures
of HBB in AGB stars in clusters of known turnoff masses and metallicity.

Overshooting (when included) is addressed as an exponentially decreasing
diffusion {\it above} formally convective regions. It makes convective cores
during the main sequence to grow larger, and also starting masses and
luminosities in AGB are then larger. However, also preliminary results
obtained when allowing displacement of convective elements {\it below}
convective regions in AGB are shown. In the ``reference" case (6\Msun), we
find that overshooting from below the convective envelope {\it totally
suppresses thermal pulses} and ultimately leads to the formation of massive
($\sim 1M_\odot$) white dwarfs rich in Carbon and Oxygen immediately below
the photosphere.

  \keywords{stars: evolution of -- AGB and post AGB -- mass loss}

\end{abstract}

\section{Introduction}

Lithium has been always the subject of extensive investigations, mainly due
to its cosmological importance. The hottest (and most massive) among the
population II main sequence stars display a plateau in the lithium abundance
(Spite \& Spite 1993), clustering around $\log \epsilon (^7Li)=2.24 \pm
0.012$ (Bonifacio \& Molaro 1997), where $\log \epsilon (^7Li)=\log
(^7Li/H)+12$. Stars in young open clusters (Pleiades, Praesepe etc.) display
an analogous plateau, with an abundance roughly a factor of ten larger
(Soderblom et al. 1990, 1993a,b, Balachandran et al. 1988, 1996). A {\it bona
fide} (and Ockham--like) explanation of this dychotomy would require the
existence of stars producing and recycling lithium in the interstellar
medium. The discovery of a few tens of lithium rich AGB stars in the
Magellanic Clouds (Smith \& Lambert 1989,1990), with $\log \epsilon (^7Li)$\
up to $\sim 3.8$, strongly recommends to take into account massive AGB stars
as possible candidates (D'Antona \& Matteucci 1991), although they might not
be the only indicted (Matteucci et al. 1995).

A mechanism for lithium production in stars has been suggested by Cameron \&
Fowler (1971) who hypothesized that, if the temperature at the base of the
envelope of an AGB star is $T_{bce} \geq 4\cdot 10^7$K, ``Hot Bottom
Burning" with enhanced production of $^7Be$ will {\bf occur}. This latter
element, if fastly carried away by convection from hig--T regions, might
decay into lithium in the outermost envelope of the star.

Static envelope models have been used to describe in a first approximation
the above mechanism, both by Sackmann et al. 1974 and Scalo et al. 1975. More
recently, Sackmann \& Boothroyd (1992) built up evolutive stellar models
through the thermal pulse (TP) phase, also testing the effect of mass loss
and metallicity. Li--production in intermediate mass stars was found within
an MLT framework, when tuning $\alpha= l/H_p > 2$. However, the present MLT
tuning of the solar model --with updated physical inputs-- ranges between
$\alpha \sim 1.5 \div 1.7$; with these low values of $\alpha$\ the onset of
HBB is shifted to larger initial masses. Of course, nobody claims that
the solar tuning of $\alpha$\ must rank as a universal constant, but the
above findings strongly recommend to test more updated and less tuning
dependent convective models.

Exploration of HBB phases according to a more modern treatment of turbulent
convection (D'Antona \& Mazzitelli 1996) started after the availability of
the FST model by (Canuto \& Mazzitelli 1991,1992). HBB in solar metallicity
stars of intermediate mass was found to be a straightforward and
tuning--independent consequence of use of the FST model. In the present
follow--up of the '96 work, the version of the ATON code ({\bf ATON 2.0},
Ventura et al. 1998) has been implemented with a diffusive algorythm for
chemical evolution which can fully address the problem of lithium production
in AGB stars. We delay to a following paper the tuning of overshooting and
mass loss required to fit the observations of AGB Lithium rich stars in the
Magellanic Clouds. Here we test the role of different input parameters.

 \begin{table}
  \caption[]{Inputs for the computed models}
	\medskip
   \begin{tabular*}{8.8cm}{ccccccc}   \hline
	\medskip
	$M$ & Start & Conv. & $\zeta$
	& $\dot M$ ($\eta$) & Z & $M_c/M_{\odot}^a$
	 \\
	   \hline
	  \medskip
	 6 & PMS & FST & 0 & 0 & 0.02 & 0.931 \\
	 6 & $D=10^{-5}$ & FST & 0 & 0 & 0.02 & 0.929 \\
	 6 & MS & FST & 0 & 0 & 0.02 & 0.931 \\
	 6 & PMS & MLT & 0 & 0 & 0.02 & 0.931 \\
	 5 & PMS & FST & 0 & 0.1 & 0.02 & 0.925 \\
	 5 & PMS & FST & 0 & 0.05 & 0.02 & 0.925 \\
	 6 & PMS & FST & 0 & 0 & 0.01 & 0.945 \\
	 6 & PMS & FST & 0 & 0 & 0.001 & 0.988 \\
	 6 & PMS & FST & 0 & 0.3 & 0.02 & 0.922 \\
	 6 & PMS & FST & 0.02 & 0.1 & 0.02 & 0.98$^b$ \\
	   &     &     &(symm)&  &      & \\
	 6 & PMS & FST & 0.02 & 0.1 & 0.02 & 1.013 \\
       5.5 & PMS & FST & 0.02 & 0.1 & 0.02 & 0.962 \\
	 5 & PMS & FST & 0.02 & 0.1 & 0.02 & 0.925 \\
       4.5 & PMS & FST & 0.02 & 0.1 & 0.02 & 0.874 \\
	 4 & PMS & FST & 0.02 & 0.1 & 0.02 & 0.812 \\
       3.5 & PMS & FST & 0.02 & 0.1 & 0.02 & 0.665 \\
	    \hline
   \end{tabular*}
\begin{list}{}{}
\item[$^{\mathrm{a}}$] Value of the
core masses refer to the interpulse phase between the fifth and
 the sixth pulse.
\item[$^{\mathrm{b}}$] Core mass at the end of computation.
\end{list}
 \end{table}

Our standard model will be a $6M_{\odot}$, $Z=Z_\odot=0.02$, with an initial
$D$--abundance (by mass) $X(D)=2\cdot 10^{-5}$. We investigate the effects of
different convection models, initial deuterium abundances, metallicity,
overshooting, mass loss rate and initial stellar mass on the AGB phase.
Mainly, we focus our attention on the core mass versus luminosity relation,
on the amount of lithium produced, and on the duration of the phase during
which the star can be seen as lithium rich. Figure 1 shows the HR
diagram of the main computed set of tracks from 3.5\Msun\ to 6\Msun. The
tracks shown include mass loss and core overshooting. The Thermal Pulse
phase, shown only for the masses 4, 5 and 6\Msun, appears as a series of
loops in the diagram at luminosities between $\log L/L_\odot \simeq 4.4$\ and
4.8.

\begin{figure}
\caption[]{HR diagram of 3.5, 4, 4.5, 5, 5.5 and 6\Msun models of solar
chemistry evolved from the pre--main sequence to the first thermal pulses.}
\label{hrtot}
\end{figure}

\section{Stellar modelling}

The main features of the {\bf ATON 2.0} code are described in details in
Ventura et al. (1998). In this paper we rediscuss the coupling between
nuclear and turbulent chemical evolution (2.1). We also briefly recall
the main micro/macrophysical inputs:
\begin{itemize}
\item{Radiative opacities from Rogers \& Iglesias (1993), and from
Alexander \& Ferguson (1994) for the low--temperature regime;}
\item{neutrino losses from Itoh et al. (1992);}
\item{electron conduction from Itoh \& Kohyama (1993);}
\item{OPAL EOS (Rogers et al. 1996) for $3.7 < \log T < 8.7$ is adopted. In
the low--T -- high--$\rho$ limit, the Mihalas et al. (1988) EOS is also
included;}
\item{the nuclear network explicitly accounts for the 14 elements: $^1H$,
$^2D$, $^3He$, ${\bf ^4He}$, $^7Li$, $^7Be$, $^{12}C$, $^{13}C$, $^{14}N$, $^{15}N$,
$^{16}O$, $^{17}O$, $^{18}O$, $^{22}Ne$. 22 reactions are considered, from
the light elements burning up to $^{12}C$ ignition. Cross--sections are taken
from Caughlan \& Fowler (1988) and screening factors by Graboske et al.
(1973). Although in HBB very large temperatures can be met in the H--burning
shell, we omitted the Na--Al cycle since its contribution to the total energy
generation is always negligible. More about this subject in the following;}
\item{the size of convective regions is evaluated according to the
Schwarzschild criterion. The convective fluxes can be computed either by the
FST model (Canuto et al. 1996, updating the Kolmogorov constant to 1.7), or
by the MLT (Vitense 1953). In these latter models, we use our solar tuning of
the free parameter $\rm \alpha=l/H_p=1.55$.}

\end{itemize}

\subsection{Chemical evolution and overshooting}

Instead of addressing chemical mixing inside convective regions within the
instantaneous mixing approximation, we adopt the more physically sound
full--coupling between nuclear evolution and turbulent diffusion (see Ventura
et al. 1998). This is necessary both to follow lithium
production/destruction, and to allow for a self--consistent treatment of
diffusive overshooting in conditions when leakadge of an high--T
CNO--burning shell is present. Let us discuss
this latter point.

In HBB conditions, even in the absence of oveshooting from {\it below} the
convective envelope, convection penetrates the H--burning shell such that a
non negligible fraction ($\sim 20 \div 30 \%$) of the star's luminosity is
generated inside a turbulently mixed region (see later). Of course, this
effect is increased if we allow for some amount of overshooting.

Our results show that the following two problems arise:

\begin{itemize}

\item{the nuclear time scales of some CNO reactions can become of the same
order of magnitude of the convective time scales, and}

\item{the local H/He production/destruction can be {\it relatively} large
during one single time step.}

\end{itemize}

The solution of the second problem necessarily leads to the requirement of
using a larger--than--zero order for the numerical integration of the nuclear
evolution. We adopt the scheme by Arnett \& Truran (1969) which, being an
implicit, first order treatment, gives integration errors $\propto (\delta
X)^2$ instead of $\propto (\delta X)$ (remember that $\delta X \ll 1$).

The above procedure requires however the explicitation of {\it all} the
chemical abundances not only {\it before}, but also {\it after} the time
step, including the effect of mixing. So, if we want to account also for the
first problem above (similiarity of time scales), no semi--explicit solution,
point by point and element by element along the structure (as for instance in
Herwig et al. 1997) is allowed. The only fully self--consistent solution is
to store {\it all the matrix elements along the whole structure}, and invert
the general matrix. This is a burdensome task, and also explains why we limit
to 14 elements our network. Our chemical evolution scheme is however more
physically sound (actually: the only physically sound one) than any other
used up today, if extreme cases of HBB are to be considered.

As for overshooting, in our code we can follow two distinct approaches:

\begin{itemize}

\item {In case of tests with the oversimplified approximation of
instantaneous mixing, we simply force full mixing up to a fixed distance from
the formal convective border. The extent of the overshooting region, usually
taken as a fraction of $H_p$, is a free parameter.}

\item {In case of diffusive mixing, we use the formalism described by
Ventura et al. (1998), allowing an exponential decay of the convective
velocity out of any convective border of the form:
\begin{equation}
u=u_b\cdot exp ({1\over \zeta f_{thick}}ln{P\over P_b}) \,,
\label{oversh}
\end{equation}
where $u_b$ and $P_b$ are turbulent velocity and total pressure at the
convective boundary, P is the local pressure, $\zeta$ is a free parameter.}

\end{itemize}

Our treatment of diffusive overshooting is then not far different from that
adopted by Herwig et al. (1997). We prefer to use pressure instead of
distance to compute exponential decay, since the first non--local models by
Xiong (1985) suggest pressure as the more correct choice, but until
overshooting occurs along short scales, pressure and distance are almost
linear with each other. The main difference is rather that Herwig et al.
(1997) use MLT convective velocities and scale lengths to evaluate the
diffusive coefficient. While FST and MLT turbulent velocities turn out to be
quite similar, it is not so for the scale lengths, especially close to the
convective boundaries (and then also close to the CNO--burning shell), where
the MLT can overestimate the local scale length by more than one order of
magnitude.



\begin{figure}
\psfig{figure=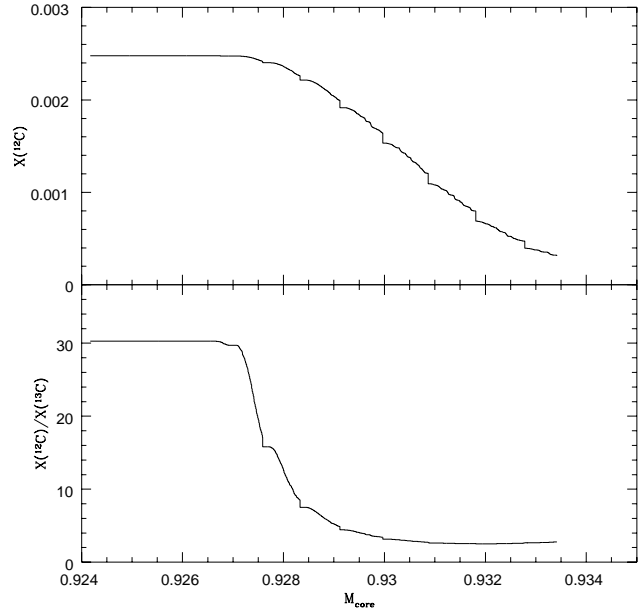,width=8.8cm}
\caption[]{Variation with core mass of $^{12}C$ surface
abundance ({\bf top})
and of the ratio $^{12}C/^{13}C$ ({\bf bottom}) within the ``standard"
$6M_{\odot}$ model. The rapid drop of $^{12}C/^{13}C$ is a clear evidence of
HBB at the base of the convective envelope.}
\label{fig_3}
\end{figure}

\begin{figure*}
\caption[]{{\bf Left}: Variation with core mass of the luminosity ({\bf
top})
and temperature at the base of the convective envelope ({\bf Bottom}) for
our standard model of 6\Msun (full lines) and for the corresponding track
assuming the MLT model for turbulent convection.
The FST model leads to larger temperatures, thus favouring
lithium production.
{\bf Right}: Variation with core mass of the surface lithium
within a $6M_{\odot}$ with two different models for turbulent convection.
{\bf Bottom}: Variation with core mass of $^3He$ surface abundance. We
note that in the MLT case the $^3He$ depletion remains negligible despite
lithium production.}
\label{fig_1}
\end{figure*}

The free parameter $\zeta$ is a measure of the size of the overshooting
region; lower $\zeta$'s allow a steeper decay of velocity and a narrower
overshooting region. Qualitatively, it is similar Herwig et al.'s free
parameter $f$. The quantity $f_{thick}$ in eq.1 is the thickness of the
convective region in fractions of $H_p$ (up to a maximum of 1), scaling the
overshooting distance according to the width of the convective zone.

The tuning of $\zeta$ has been performed by Ventura et al. (1998). For main
sequence stars of any mass (including the solar one), $\zeta = 0.02$ gives
results consistent with the analysis of open clusters HR diagrams by Maeder
\& Meynet (1991) and Stothers \& Chin (1992). We therefore assume that, at
least in a first approximation, the above tuning of $\zeta$ be representative
of the overall characteristics of overshooting, at least from the top of a
deep convective region.

As for lithium production/destruction, the main reactions are listed below,
together with the temperature at which they become important in HBB
conditions.
\begin{eqnarray}
Reaction~1: & ^7Li + p \longrightarrow 2\alpha & T\simgt 2\cdot 10^7K\\
Reaction~2: & ^7Be \longrightarrow ^7Li + e &  \tau = 53.3 days \\
Reaction~3: & ^3He+^4He \longrightarrow ^7Be~~~  &  T\simgt 4\cdot 10^7K \\
Reaction~4: & ^7Be + p \longrightarrow 2\alpha & T\simgt 8\cdot 10^7K
\end{eqnarray}
Note that the temperatures are $\sim$one order of magnitude larger than
those at which the same reactions are active in previous evolutionary
phases. This is due to the very short time scale of the whole AGB phase,
requiring extremely large reaction rates to lead to appreciable chemical
evolution.

Let us shortly resume the Li--production mechanism. Lithium is destroyed very
quickly due to proton capture (reaction 1), and the only production channel
is via Beryllium decay on the timescale $\tau$\ of reaction 2. In turn,
$^7Be$ is produced through reaction 3 and, at large temperatures, it can be
also destroyed by a proton capture (reaction 4).

The problem in getting lithium production stems from the facts that
lithium is immediately destroyed via reaction (1), and that also the ignition
of reaction (4) could prevent beryllium from decaying into lithium. The
solution suggested by Cameron \& Fowler (1971) is based on two facts:
\begin{itemize}
\item {for temperatures $T<8\cdot 10^7 K$, the rate of reaction (4) is not
so large to prevent beryllium decay: moreover, reaction (4) is
slower than mixing at the base of the convective envelope;}
\item {inside the very expanded envelope of AGB stars, beryllium decay is
slow enough that lithium production occurs in cooler regions of the envelope
of the star, where it can survive to show up at the surface.}
\end{itemize}

In this scheme, nuclear evolution and mixing are so strictly intertwined
that they cannot be deat within the instantaneous mixing approximation.
Surface lithium would be in fact mixed at once, to be instantaneously
destroyed at the base of the envelope.

Of course, the solution by Cameron \& Fowler (1971) can work if reaction (4)
is not running too fast at the base of the external envelope (Sackmann \&
Boothroyd 1992, 1995), otherwise proton capture would leave no beryllium
available to decay into lithium. Computations show that, for $T<8\cdot 10^7$
reaction (4) is slower than mixing and overabundances of lithium can be
observed. If HBB occurs at larger temperatures, beryllium burns at the base
of the envelope, and lithium in the envelope begins to decrease.

\section{Evolution of a $6M_{\odot}$ star}

We first discuss, for reference, the evolution of a $6M_{\odot}$ model
starting from the pre-main sequence (PMS), with no mass loss, no overshooting
and FST convection. The chemistry is solar ($Y=0.28, Z=0.02$), and initial
deuterium and lithium abundances (in mass) of respectively $2\cdot 10^{-5}$
and $2\cdot 10^{-8}$ ($\log \epsilon (^7Li)= 3.6$) are assumed. The track has
been followed until the star underwent a handful of Thermal Pulses (TPs). We
shortly elucidate the surface chemical evolution, mainly of $^3He$ and
$^7Li$, prior to the beginning of the AGB phase. Remember that $^3He$ is a
key ingredient (reaction {\bf 4}) in $^7Li$--production.

\begin{figure}
\psfig{figure=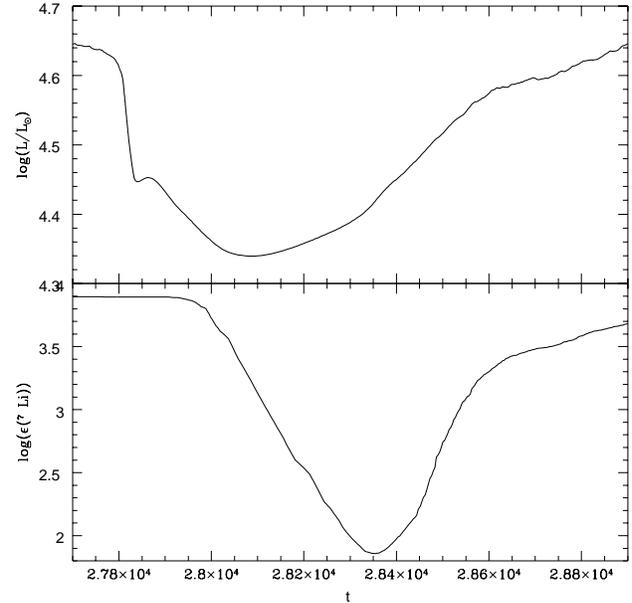,width=8.8cm}
\caption[]{Variation of luminosity and surface lithium abundance during a
thermal pulse. Times were normalized to the 2nd dredge -- up. We note a
temporal delay between lithium depletion with
respect to the drop in luminosity, which is due to the time needed by
convection to carry matter depleted in lithium at the surface of the star.}
\label{fig_4}
\end{figure}

During PMS, D--burning leads to an $^3He$ abundance of $3\cdot 10^{-5}$ in a
large fraction of the structure. Lithium depletion is large in the interior,
but negligible at the surface, since the temperature at the bottom of the
(thin) convective envelope is always $T_{bce} < 2 \cdot 10^6$K.

\begin{figure*}
\centerline{\hbox{
\psfig{figure=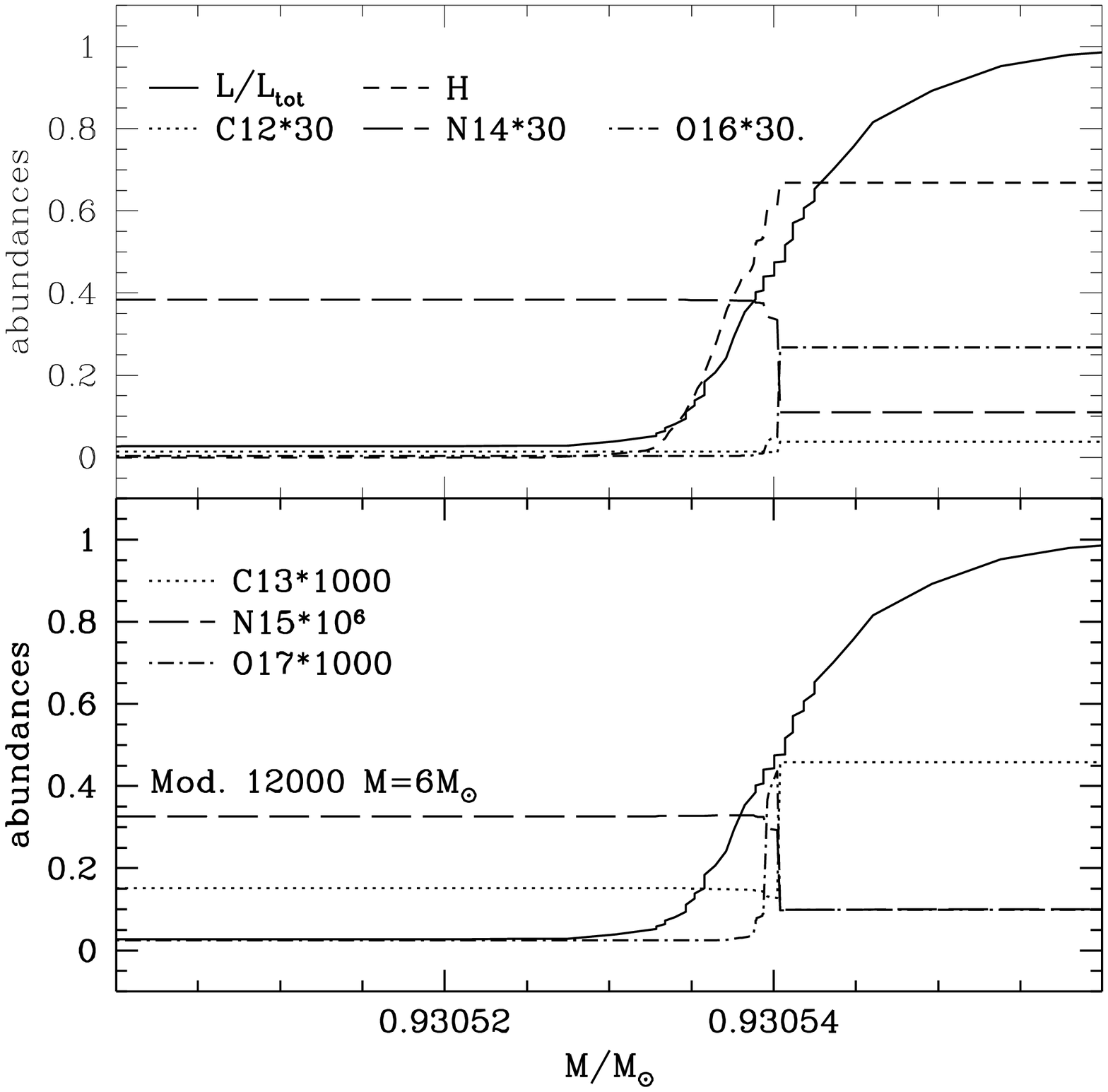,width=8.cm}
\psfig{figure=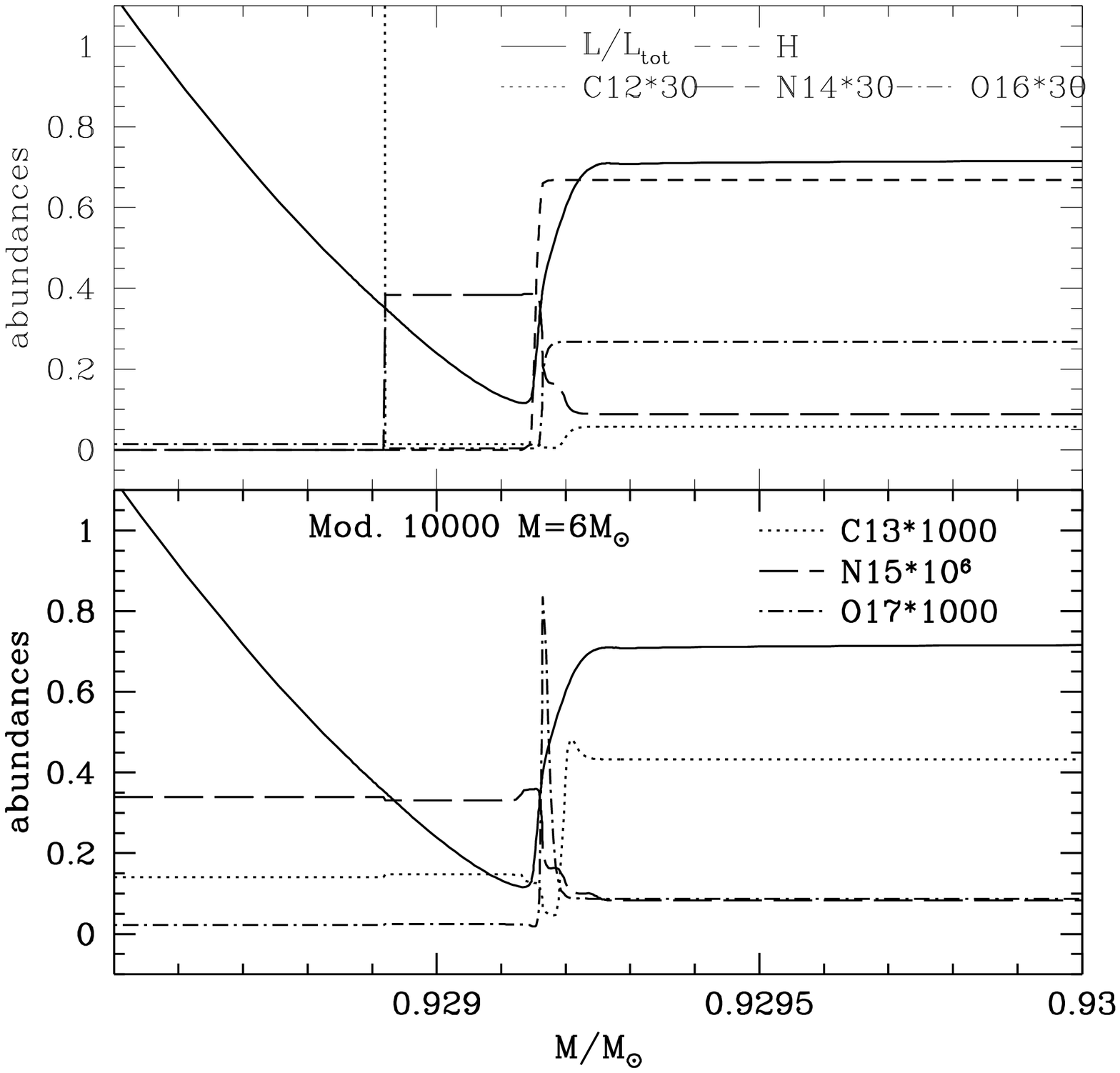,width=8.cm}}}
\caption[]{Chemical profiles in the {\it convective} hydrogen burning shell
are shown for standard models without overshooting in two different phases.
Model 12000 (left) is during stationary H--burning; model 10000 (right)
during a TP. In the interpulse phase, $\sim 30\%$ of the nuclear luminosity
is born in the lower part of the convective envelope; correct coupling
between nuclear evolution and turbulent mixing is then necessary to account
for the CNO equilibria and the energy output.}
\label{chim6base}
\end{figure*}

A first drop in lithium abundance follows the first dredge up, when the base
of the convective envelope reaches layers where lithium had been previously
destroyed. At the end of this phase, $\log \epsilon (^7 Li) = 2.1$ and the
$^3He$ abundance increases up to $6\cdot 10^{-5}$, due to the sinking of
convection down to regions where incomplete $p-p$ burning occurred. Also
$^4He$ increases up to 0.29, and the CNO isotopic abundances keep track of
mixing up to the stellar surface of CNO processed matter.

At the second dredge up, following He--ignition in a thick shell around the
C/O core and exhaustion of the H-burning shell, the $^4He$\ abundance rises
to 0.31, but no other important modifications in the isotopic abundances are
found. The lithium surface abundances starts changing significantly soon
after exhaustion of the thick He--burning shell and re--ignition of the
H--shell, when the $T_{bce}$ starts increasing and, as soon as this latter
overcomes $10^7 K$, it drops to $\log \epsilon (^7 Li) \sim 0.3$. Only when
$T_{bce} \sim 3\cdot 10^7 K$, the Cameron-Fowler mechanism begins to be
efficient, and lithium production takes over lithium burning (Fig.2
and 3). This is also the phase when the first TPs are
ignited. Since these latter are responsible for large excursions of
temperature at the bottom of the convective envelope, they also affect
lithium evolution, as detailed in the following.

The core mass in the TP phase is $\sim 0.93 M_\odot$; lower than found in
D'Antona \& Mazzitelli (1996) mainly due to the different evolutionary
scheme, and in close agreement with Wagenhuber \& Groenewegen (1998).

During the TP phase, $\log \epsilon (^7 Li)$ fastly increases up to a maximum
value of $\sim 3.8$ soon after the second major pulse. It fastly drops
shortly after the peak of each TP, due to the decrease of $T_{bce}$ following
the He--outburst, and then recovers its maximum abundance. Both drop and
recovering are slightly delayed with respect to the answer of the surface of
the star to TPs (fig.4), due to the relatively long time required
by convection to fully mix the very expanded envelope.

Going ahead with the TPs, the plateau value of $\epsilon (^7 Li)$ begins to
slowly decrease. This is due to the onset of two mechanisms:

\begin{itemize}

\item{the temperature at the bottom of the convective envelope goes on
increasing (Fig.3, bottom left panel) until reaction 4
begins to compete with reaction 2, and some $^7Be$ is no more
available to produce $^7Li$, and}

\item{in the absence of reactions producing $^3He$, this latter element
begins to be consumed by reaction {\bf 4} (Fig.3, bottom right panel),
and also $^7Be$--production is no more very efficient.}

\end{itemize}

Another mechanism worth discussing is the relatively large fraction of the
H--burning shell interested by convection. From Fig.5 we see
that a non negligible fraction of the nuclear luminosity is generated {\it
inside} the convective envelope. This explains the very fast drop in the
surface $^{12}C$ abundance, and the even faster reaching of the
$^{12}C/^{13}C$ equilibrium ratio in a very few pulses (fig.2).
Fig.5 shows the $^{15}N$ profile close to the bottom of the
convective envelope, during an interpulse. Since $^{15}N$ is a bottleneck in
the CNO cycle, any evolutive scheme other than full coupling between mixing
and nuclear evolution is bound to fail in correctly describing the nuclear
output in the convective fraction of the H--shell. This will be more and more
relevant when also overshooting from {\it below} the envelope will be
accounted for.

In conclusion, we find that a straightforward application of the FST model
immediately and nicely leads to the occurrence of HBB in AGB of intermediate
mass stars, and to the appearence of $^7Li$ overabundances, without requiring
a non--solar tuning of the model itself. Let us then start discussing the
results found when changing either the tuning of the FST model, or the
convective model as a whole.

\section{Influence of the convective model}

The FST convective model allows for a limited amount of tuning thanks to the
$\beta$ parameter (for a complete discussion and for the solar tuning see
Ventura et al. 1998). Since the standard evolution was computed with
$\beta=0.1$, we tested one more evolution with $\beta=0.05$.

From a physical point of view, this latter choice leads to a shorter
turbulent scale length close to the convective boundaries, making convection
slightly less efficient. In AGB conditions, the final result is that the base
of the envelope is, on the average, cooler. As a consequence, the
Cameron-Fowler mechanism runs at a slower pace, and the amount of lithium
produced is lower. However, some (small) differences are found only during
the first three TPs, after which the plateau value of $\epsilon (^7 Li)$
becomes insensitive to the tuning of the FST model. We found only one clear
difference between the two cases: at the peak of the TP, lithium destruction
is larger the larger is $\beta$. This is, in our opinion, a negligible
effect, since these phases are very short lived compared with the interpulse
times. We then conclude that, at variance with what happens in the MLT case,
lithium production in AGB is largely independent of tuning within the FST
convective model.

As for the tests with the MLT, we first recall that D'Antona \& Mazzitelli
(1996) examined in detail the dependence of lithium production  during the
AGB phase on the convective model. In the case of a $5M_{\odot}$ star they
found that, in the MLT case, HBB could be achieved only with a value of the
free parameter $\alpha\sim2.5$, consistent with Sackmann and Boothroyd 1992.
When assuming the solar tuning $\alpha=1.5$, no trace of $^3He$ depletion
(see their Fig.4) for the first six pulses was found (full
coupling between nuclear and turbulent evolution was not yet implemented in
the code), leading the authors to conclude that no significant lithium
production might be accomplished in that case. Here we examine the case of a
$6M_{\odot}$, for which we follow the evolution with MLT and the present
solar tuning $\alpha=1.55$.

The results are shown in Fig.3 where we report the evolution vs.
core mass of surface luminosity, $T_{bce}$, $\epsilon (^7Li)$ and $X(^3He)$
in the FST (standard) and  MLT sequences. First note that, before the onset
of the TPs, the two tracks show quite different surface lithium abundances,
the MLT models maintaining $\sim$100 times more $^7Li$ than the FST one.

A relevant result is that the core mass -- luminosity relation is largely
affected by the convection model. At core mass $0.93M_{\odot}$, FST models
are $\sim$2 times more luminous than MLT ones. This is due to the
intrinsically lower efficiency of the MLT convection with respect to the FST
one, and we have already seen that the lower is the efficiency, the lower is
$T_{bce}$. Since the temperature in the H--burning shell is quite close to
$T_{bce}$ (there is a merge between convective envelope and shell), solarly
tuned MLT models are cooler and less luminous, and can be made consistent
with FST ones only at the price of artificially enhancing the convective
efficiency by increasing $\alpha$.

At the same time, the lower $T_{bce}$ in the MLT case makes the
Cameron-Fowler mechanism less efficient and, in fact, MLT $^7Li$--production
takes longer ($\sim$10 TPs) before being comparable to the FST one. Also
$^3He$ begins to be depleted, in the MLT case, only after a number of pulses,
when $^7Li$ approaches the maximum, while in the FST case it starts
decreasing at the first pulses.

We did not repeate the computations with larger values of $\alpha$, since the
present models are largely consistent with those by D'Antona \& Mazzitelli
(1996). The preliminary conclusion which can be drawn by the above results is
then that also MLT models with solarly tuned $\alpha$\ can give rise to
Li--rich AGB's, but the range of initial masses undergoing HBB is very
narrow; below 6\Msun\ no Li--enhancement can be obtained, and already at $6.5
\div 7.0 M_\odot$ degenerate $^{12}C$ ignition in the core stops evolution
before the onset of TPs. In the FST case, as we will see later, already stars
of $4.0 \div 4.5 M_\odot$ can attain HBB.

These conclusions will be later strengthened, when discussing the occurrence
of (possible) overshooting from the core, and (unavoidable) mass loss. As we
will show, the chances are that these stars undergo a relatively low number
of TPs before planetary nebula ejection. Little or no Li--enrichment of the
interstellar medium is then expected if the Cameron-Fowler mechanism does not
{\bf occur} very early in AGB.

\section{Overshooting "from above"}

A small amount of convective overshooting at least from the core, both in MS
and during the central He--burning phase, is likely to {\bf occur} during the
evolution of intermediate mass stars. In Ventura et al. (1998), the diffusive
scheme here adopted to account for overshooting has been tuned as discussed
in Sect. 2.1.

\begin{figure*}
\caption[]{Comparison between the standard 6\Msun\ evolution (dotted line)
and the model including core overshooting and mass loss (solid line).
Overshooting leads to a larger core mass and faster evolution.}
\label{6mbov}
\end{figure*}

More disputable is the existence of overshooting from {\it below} a
convective region. The helioseismological results by Basu (1997)
severely constrain the amplitude of the region below the solar convective
envelope, in which the temperature gradient is adiabatic. However, since
non--local computations (Xiong 1985) show that the overshooting
gradient sticks to the radiative one, we still get no informations about the
amplitude of the {\it chemically mixed} region (if any), below a convective
shell.

For this reason, we separated the two cases: overshooting {\it from above} a
convective region, and {\it symmetric} overshooting. In the first case,
discussed here, diffusive overshooting (fully coupled to nuclear evolution)
is allowed from the top of {\it any} convective region, including the
convective He--C shell at the peak of a TP, with the same tuning $\zeta =
0.02$. In the second case, discussed in a later section, the same tuning is
applied to overshooting also from {\it below} any convective shell,
including the external envelope and the He--C shell at the TP.
In the following of this section, then, we will implicitly assume that
overshooting occurs only from the top of {\it any} convective zone.

The main difference found with respect to the reference case (no
overshooting) is the increased size of the H--exhausted core at the end of
the MS phase. In fact, during central He--burning, the larger opacities of C
and O lead to a "semiconvective--like" condition largely resetting the effect
of a small amount of overshooting. Lastly, during TPs, the entropy barrier
is large enough that even overshooting from the top of the He--C rich shell
does not lead to mixing with the external H--rich layers.

At the end of the second dredge--up phase, then, the H--exhausted core mass
is larger than without overshooting by $\sim 0.08 M_\odot$, that is: $\sim
1.01 M_\odot$. This is to be compared to the $\sim 1.05 M_\odot$ required, in
this phase, to ignite $^{12}C + ^{12}C$ degenerate burning off--centre,
leading to sudden stop of the AGB (and to the birth of an O--Ne--Mg--rich
WD?). If we define $M_{up}$ as the initial stellar mass for which no AGB
evolution occurs, even a small amount of overshooting in MS leads to $M_{up}
\sim 6.5 M_\odot$.

The $6.0 M_\odot$ star, however, can still proceed through the TP phase. The
larger core mass leads to a faster increase of $T_{bce}$ than in the
reference case, as soon as the H--burning shell is re--ignited. HBB takes
over even prior to the ignition of TPs, and the surface lithium abundance
reaches a maximum value $\log \epsilon (^7Li) \sim 4.1$. Also the
$^{12}C/^{13}C$ surface ratio fastly drops to the equilibrium value.

These latter features are to be kept into account when comparing theory with
observations. In {\it any} MLT environment, irrespective of the tuning of
$\alpha$, it is in fact very hard (overshooting or not) to get HBB with the
corresponding surface chemical signatures before the ignition of TPs. If, as
expected according to the observations of relatively low--luminosity
C--stars, the third dredge--up mechanism begins occurring already at the
first TPs, s--process elements should almost always accompany
Li--overabundances, as it is happens in the Magellanic Cloud stars (Smith
et al. 1995). If HBB takes over prior to the ignition of TPs, as in the
present case, we can certainly expect to observe Li--overabundances without
signature of s--process elements, as it has been recently found in a sample
of galactic massive AGBs by Garcia Lario et al. (1998).

Of course, this result is insufficient to discriminate between convective
models for the simple reason that, although we {\it reasonably} expect third
dredge--up to immediately follow the onset of the TP phase, no convincing
theoretical models still exist showing that this is indeed the occurrence,
and, further, the lack of s--process enhancement in Garcia Lario et al.
sample may have several other convincing explanations (see later). We only
want to stress that, with FTP convection and a small (almost unavoidable,
perhaps) amount of overshooting in MS, we spontaneously find
Li--overabundances at the beginning of the AGB, prior to the ignition of
TPs.

In Fig.6 we compare luminosity, $T_{bce}$\ and lithium evolution
in the reference model and in a model with overshooting and mass loss (see
next section for this latter mechanism). We see a fast decrease of the
surface luminosity after $\sim 10^4$yr from the second dredge up,
due to the large mass loss rate adopted. Further, the evolution is faster,
consistent with the higher core mass, and $T_{bce}$\ and lithium are larger,
with respect to the no--overshooting case.

\section{Mass loss}

Mass loss in AGB is another physical feature to be obviously accounted for.
Among the various possible choices, we decided to follow Bl\"ocker's (1995)
recipe, to assume a Reimers' rate:

\begin{equation}
\dot M_R=4\cdot 10^{-13}\eta \cdot {LR\over M}
\end{equation}

until the Mira oscillation period is lower than $\sim 100$ days, to switch
to a much larger rate:

\begin{equation}
\dot M= \dot M_R \cdot L^{2.7}/M^{2.1}
\label{bloc}
\end{equation}

when in AGB. The semiempirical tuning of the free parameter $\eta$ according
to Bl\"ocker's models is around 0.1, and we adopted the same value. However,
Bl\"ocker's (1995) core masses in AGB are among the lowest available in the
literature (Wagenhuber \& Groenevegen 1998). Also his luminosities are then
lower than the present ones for the same initial masses and evolutionary
phases, and the stiff dependence of $\dot M$ on L makes our {\it initial}
mass loss rates in AGB $\sim 5 \div 10$ times larger than due. This is not
yet relevant for the purposes of the present paper, since we are here
interested in semi--quantitative comparisons only. We made some test
computation on a 5\Msun\ track, computing it with $\eta=0.1$\ and with
$\eta=0.05$. The comparison of the results is shown in Fig.7:
the lower mass loss rate leads to a longer run of TPs, and the lithium
abundance at the surface remains large for the whole run. In general, we can
say that extensive computation of grids of evolutionary tracks will require
values of $\eta$\ smaller than 0.1 for our models. Even more so, if
overshooting (and corresponding increase of core mass and luminosity) is
considered. The following discussion will however show that there is no
linear relation between the value of $\eta$ and the various evolutionary
occurrences.

\begin{figure}
\caption[]{Variation with time of $^7Li$, luminosity, $T_{bce}$ within two
$5M_{\odot}$ models computed by assuming two different mass loss rates
($\eta=0.1$\ and $\eta=0.05$). We note that increasing $\dot M$ leads to a
more rapid decrease of the luminosity of the star.}
\label{fig_9}
\end{figure}

Mass loss dramatically modifies the evolution of our models. For the $6
M_\odot$ case, after a few $(\sim 5)$ pulses the envelope mass decreases
below $1 M_\odot$, stopping HBB. After one--two further pulses, planetary
nebula ejection should be expected. The duration of the lithium--rich phase
is consistently reduced, the more the larger is the initial mass of the star.
The effect is then relevant also to correctly determine the IMF of
lithium--rich AGB stars.

Another feature to be stressed is that mass loss leads to early deviations
from any core mass--luminosity relation. In fact, surface luminosity starts
decreasing very soon. This affects the appearence of lithium rich stars in
the HR diagram, since the same Li-abundance one would have found with no mass
loss is now achieved at a lower luminosity.
This is shown in Fig.8 where the AGB evolution of the 6\Msun\
track without mass loss and with $\eta =0.3$\ are compared.

\begin{figure}
\caption[]{Comparison between the 6\Msun\ evolution without mass loss
(track at the left) and the same evolution including mass loss
with $\eta=0.3$.} \label{agbml} \end{figure}

There is however another relevant consequence of the effect of mass loss
on surface luminosity. It acts as a sort of feedback on mass loss itself.
Namely, a lower value of $\eta$ would have allowed a slightly larger growth
of the luminosity after which, due to the stiff dependence of $\dot M$ on L,
almost identical conditions as those shown here would have been met. Being
$\dot M \propto L^{3.7}$, a decrease of $\eta$ by a factor of ten would
only have increased by a factor $\sim 2 \div 2.5$ the total number of pulses
for the $6 M_\odot$ star. At the upper mass limits, intermediate mass stars
are not expected to go through more than $\sim 15 \div 20$ TPs with {\it
realistic} mass loss rates, and the number increases roughly $\propto M^{-2}$
down to $\sim 4 M_\odot$ which, according to the present computations, is the
lower limit to get HBB lithium overabundances for a solar chemistry (see
later).

\section{Dependence on initial deuterium and lithium}

After discussing the main effects due to overshooting and mass loss, let us
turn to the exploration of the rest of the parameters space. Reaction 3
suggests that Li--production in AGB should be sensitive to the $^3He$
abundance in the envelope. Also, in Sect. 3 we have seen that the amount of
$^3He$ at the beginning of the AGB phase somewhat depends on the residual
$^3He$ left at the end of the PMS phase due to early D--burning. It is then
interesting to check what happens of Li--production if the initial
D--abundance is changed.

We then ignored PMS D--burning, simply by running an evolution starting from
zero age MS with no $^3He$. As expected, the overall properties of the run
are almost identical to those of the reference track. A vanishingly small
difference ($<0.2\%$) in core mass at the beginning of the AGB phase is
present, due to a cumulative difference in the efficiency of $p-p$ burning at
the border of the convective core during the main sequence phase. Only in
AGB, with less $^3He$ available (Fig.9, bottom panel), the track
starting from MS begins to show a behavior of its own in terms of lithium
abundance (Fig.9, top panel). In fact, during the first pulses,
the growth of $\epsilon (^7Li)$ is slower, and the maximum value --attained
after $5\div 6$ pulses-- is $\sim 3.7$.

Since the temperature profile within the whole convective region is the same
for both the present model and the reference one, we can say that the $^3He$
abundance is the second most important parameter other than $T_{bce}$
influencing lithium production in AGB. This is confirmed by one further run
starting from PMS, with an initial D--abundance half of the standard one. As
expected, we found an intermediate situation between the two above.

\begin{figure}
\caption[]{Variation with core mass of $^7 Li$ ({\bf top})
$^3 He$ ({\bf bottom}) surface abundances along the evolution of
$6M_{\odot}$ models computed with different initial deuterium
abundances. The MS track refers to a model computed starting from the
main sequence phase. Times for the three models have been normalized
in the course of the second dredge--up.}
\label{fig_5}
\end{figure}

A comments is due to this result. For the test structure, the amount of
$^3He$ produced by incomplete $p-p$ chain during the evolutionary phases
prior to AGB is about twice the amount due to early D--burning, {\it with a
proto--solar abundance of deuterium}. If (as expected) the cosmological
D-abundance was larger than the present one (by a factor of 2 or more,
depending on the interpretation of observations of high redshift absorbers
along the lines of sight of distant quasars (Tytler et al. 1996, Songaila
et al. 1997), intermediate mass stars of earlier stellar populations
must have been more vigorous producers of $^7Li$ than the present ones.

Lastly, also the influence of initial lithium the AGB Li--production was
tested. An evolution starting with a lithium abundance of $X(^7Li)=10^{-8}$
was followed. As expected, and in full agreement with Sackmann \& Boothroyd
(1992), no effect on the final results was found. The structure loses any
memory of the previous {\bf history} of lithium at the beginning of the AGB phase,
and the lithium abundance during TPs is solely determined by $T_{bce}$ and
the $^3He$ abundance.

In summary, the results are independent of the initial lithium abundance, but
somewhat sensitive to the initial deuterium abundance, at least during the
first TPs. If (as we will see) large mass loss rates can terminate AGB
evolution shortly after the ignition of TPs, complete evolutions starting
from PMS, with D--burning included, are necessary to correctly evaluate
Li--production by intermediate mass stars.

\section{Different initial chemistries}

To fully exploit Li-production during the life of the Galaxy, it is obviously
necessary to compute also AGB models for metal abundances lower than solar.
We then followed two more evolutions of $6 M_\odot$ stars with,
respectively, Y=0.26, Z=0.01 and Y=0.24, Z=0.001. Some results are
shown in Fig.10.

The first effect found was an increase in the core mass at the beginning of
the TP phase, when decreasing Z. Most of the effect is due to the larger
convective core mass in MS, consistently with the larger MS luminosity. As a
consequence, also $T_{bce}$ is larger for lower Z. This leads to a larger
consumption of $^7Li$ prior to the AGB phase but, as we have seen before,
the evolution of lithium before the Cameron/Fowler mechanism ignites is
of no consequences for the following Li--evolution.

In AGB, due both to the larger core mass and to the lower surface opacity,
low-Z models have larger values of $T_{bce}$  and larger surface
luminosities. This accelerates Li--evolution: the maximum value of $\log
\epsilon (^7Li)$ ($\sim 4.3$ for Z=0.001) is reached in a shorter time with
respect to the solar case. Of course, also the following evolution
(depletion of $^3He$ and of $^7Li$) is accelerated, since $T_{bce}$
overcomes $10^8$K. Let us insist that, to study the evolution {\it as a
whole} (i.e.: also when ignoring the details of the surface Li--evolution)
of these structures, full coupling between nuclear and turbulent chemical
evolution is an absolute prerequisite, since the CNO isotopic concentrations
in the convective fraction of the H--burning shell largely determinate the
nuclear output.

\begin{figure}
\caption[]{Variation with time of the surface lithium abundance, luminosity
and temperature at the bottom of convection, computed
for a $6M_{\odot}$ model with three different metallicities. Lower $Z$
models burn lithium faster due to the larger temperature at the base of
the envelope.
Computations were made without including mass loss and overshooting.
}\label{fig_10}
\end{figure}

The total duration of an intermediate mass star as a super--lithium rich
object is then a strong function of the metallicity: very roughly, $\tau_{Li}
\propto Z^{1/3}$.

\begin{figure}
\caption[]{The evolution with time of the difference in mass between the
H--exhausted and the He--exhausted layers, when asymmetric (dotted line)
and symmetric (solid line) overshooting are considered. In the latter case,
although the total mass of the He--rich intershell is initially
slightly larger, overshooting from the bottom of the convective envelope
prevents accumulation of helium and the onset of TPs.}
\label{fig_mshell}
\end{figure}

\section{Overshooting "from below"}

The last test performed on the $6 M_\odot$ star before going to the
discussion of lower mass models is the case in which exponentially decaying
diffusive overshooting, fully coupled to nuclear evolution, occurs also {\it
from below} any convective shell/envelope all through stellar evolution. Also
in this case, the value of the free parameter $\zeta$ has been fixed to 0.02
({\it symmetric} overshooting).

A similar framework has been shown by Herwig et al. (1997) to produce third
dredge--up and carbon stars for relatively low initial mass structures. Our
results for initial masses $M \leq 5 M_\odot$, which will be discussed in a
next paper, are in substantial agreement. Overshooting {\it from below} leads
to the origin of C--stars due to two combined effects:

\begin{itemize}
\item during the TP, helium in the He--C--rich convective shell is
overshooted to inner regions, where the temperature is larger. Due to the
very stiff power dependence of the $3\alpha$ reactions on T, and to the fact
that a relatively large abundance of He is mixed down, the power of the
pulse at the peak is largely increased. The feedback of the structure to the
larger energy output leads to a deeper sinking of the external convective
envelope when the H--shell is turned off, and:
\item overshooting from below the envelope as a whole, obviously facilitates
the third dredge--up process.
\end{itemize}

What we examine here is instead likely to be an extreme case, which can be
however of interest in the framework of the most luminous, Li--rich AGB stars
with little (or no) trace of s--processed elements (see for instance Garcia
Lario et al. (1998) in a spectroscopic survey of IRAS sources selected to
probably contain the most massive galactic AGBs).

In Fig.12 we compare the evolution of the 6\Msun\ with core
overshooting and mass loss described in Sect. 6, with an analogous evolution
including also overshooting {\it from below} the convective regions (also
defined: {\it symmetric} overshooting case). The two evolutions have been
shifted in time such that, at the same moment, in both cases the luminosity
contribution from the {\it thick} He--burning shell following the second
dredge up drops to 20\% (Fig.6c).

In the case with {\it symmetric} overshooting, the decline of the efficiency
of the He--burning shell preceding the onset of the TP phase is faster than
in the {\it asymmetric} case, ultimately dropping to zero. Of course, in the
other case, after a minimum contribution of the He--shell, thermal pulses
begin taking place. Let us focus our attention on Fig.11 to
understand the reason for the different behaviors.

\begin{figure*}
\caption[]{Comparison between the 6\Msun\ evolution
including core overshooting and mass loss (dashed) and the evolution
including `simmetric' overshooting. In this latter case, no TPs are present
due to penetration of overshooting in the He--rich intershell.}
\label{6movsim}
\end{figure*}

The two tracks show the thickness in mass of the region between the first
grid point in which hydrogen is completely exhausted, and the last grid point
where some helium is still present. Then, they do not show the total amount
of intershell helium in solar masses. Steps down mark the instant when helium
is completely exhausted in the lower grid point in which it was formerly
present and, as can be seen, zoning in this region is relatively coarse
($\sim 0.0005 M_\odot$) since the occurrences there are almost {\it linear}.
The upward slopes following each step down show instead the effect of the
outward H--shell shift, with corresponding accumulation of fresh helium in
the shell.

The dotted line ({\it asymmetric} overshooting) shows that, following each
step downward, accumulation of new helium is always present (even if the
total thickness of the He--rich region decreases). Around $t = 1.7 \cdot
10^4$ years, the accumulation of helium leads to the ignition of the first
(still weak) TP. From the solid line ({\it symmetric} overshooting) we
instead see that, in a much earlier phase, downward steps are not followed
any more by accumulation of new helium. On the contrary, approximately when
the TPs should begin, the thickness of the He--rich region begins to drop
fast.

This is the consequence of overshooting {\it from below}. Soon after the
ignition of the H--burning shell, this latter begins being fully penetrated
by overshooting, ultimately leaking through the whole shell and into the
H--exhausted (and C--enriched) mantle below. At this point, all the helium
produced by H--burning is mixed up in the envelope, and even some helium (and
carbon) from the He--mantle are mixed up. Lack of accumulation of helium
leads to the impossibility of igniting TPs, and the star goes on stationarily
losing mass in conditions of HBB (and then of Li--overabundance, but with
little surface carbon, which is almost entirely transformed into $N^{14}$ by
the CNO cycle). Ultimately, when the He--mantle is almost completely
destroyed, the large abundances of $C^{12}$ leaked by overshooting and
convected inside the furiously CNO burning H--shell cause a fast increase in
surface luminosity (and mass loss) terminating the AGB phase.

Let us insist on the requirement of a full--coupling between mixing and
nuclear evolution, especially in these phases. More than this, the diffusion
coefficient must be {\it physically sound}, since the evolution of the
H--shell is largely sensitive to the velocity with which fresh $C^{12}$ from
below is convected outwards. Now, if there is agreement on something among
MLT users, it is clear that MLT is unable to correctly predict mixing
timescales at the base of an AGB envelope (Wagenhuber \& Groenewegen 1998).
The FST convective model, on the contrary, is not bound to unphysically
large scale length, and also predicts more sound convective velocities,
since the turbulent fluxes are far more consistent with experimental ones.
It is then unlikely that such extreme conditions could be investigated with
descriptions of turbulence and mixing different from the present ones. This
is also true of the first--order expansion of the nuclear evolution as a
whole (Arnett \& Truran), since in thin shell conditions one can easily get
(relatively) large values of $\Delta X$, and a zero--order evolutionary
scheme would badly overestimate H--consumption.

\subsection{Possible evolutionary consequences}

Figures 13 show two models in the latest phase of this evolution:
comparison shows how the hydrogen shell {\it penetrates} into the structure,
dredging up the top of the carbon core. Notice that this kind of evolution
{\it completely wipes out the helium buffer layer}. So, we ultimately expect
a (massive) white dwarf showing carbon and even oxygen in large quantities up
to the stellar surface, below a very thin H--rich layer (if any).

This is just the structure {\it ad hoc} hypotesized in hydrodynamic
theoretical models to explain the {\it fast nova} mechanism (e.g. Prialnik
\& Kovetz 1984, Kutter \& Sparks 1989), and actually seen in nova ejecta
(Williams 1985, Truran \& Livio 1986). In fact, not only large masses (i.e.
surface gravities) are required, but the short evolutionary times (hours) do
not allow the {\it complete} CNO cycle to operate, because of the bottleneck
of the $^{13}N$ decay, which takes $\sim$10 min. So, only the energetics of
the $^{12}C + p$ reaction is available, and the observed energy outputs (and
chemical ejecta) are consistent with large overabundances of $^{12}C$, about
$20 \div 40$\% by mass. In the presence of an He--intershell, hardly the
H--rich matter accreted to the surface could be able to mix with the C--O
rich core. In the present framework, instead, {\it symmetric} overshooting
and consequent shortage of the He--intershell just for the more massive
pre--WDs spontaneously leads to the conditions in which {\it fast novae}
would {\bf occur}.

This may be also the evolutionary path which finally leads to the observed
surface chemical abundances of some memebers of the PG1159 class of hot white
dwarfs, including their prototype. In fact, Werner et al. (1991) have found
significant overabundances of carbon and oxygen in their spectra, and models
of nonadiabatic pulsations of PG type stars require a large Oxygen and Carbon
overabundances just below the photospere to drive the pulsational
instabilities (Stanghellini et al. 1991).

\begin{figure*}
\centerline{\hbox{
\psfig{figure=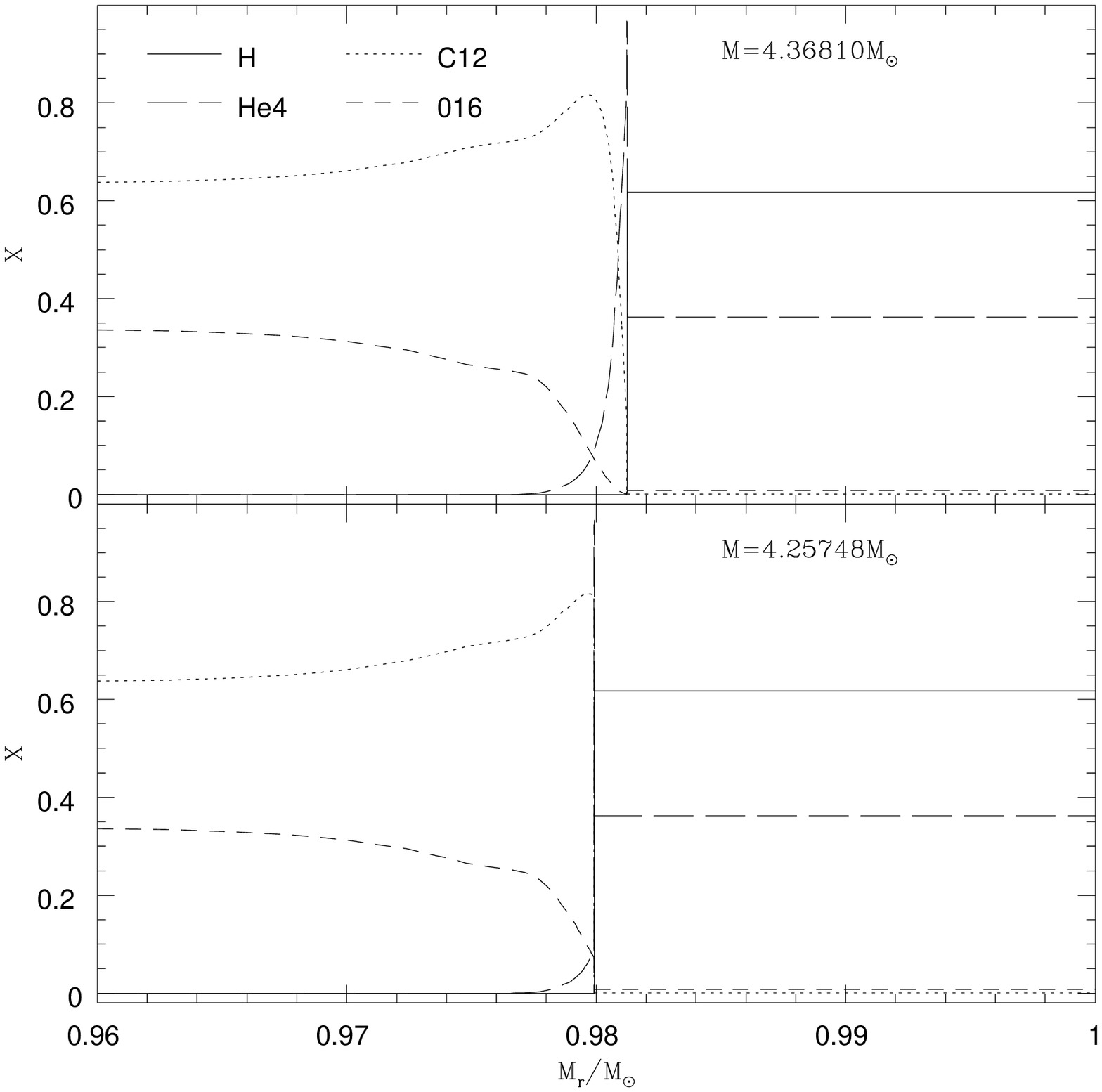,width=8cm}
\psfig{figure=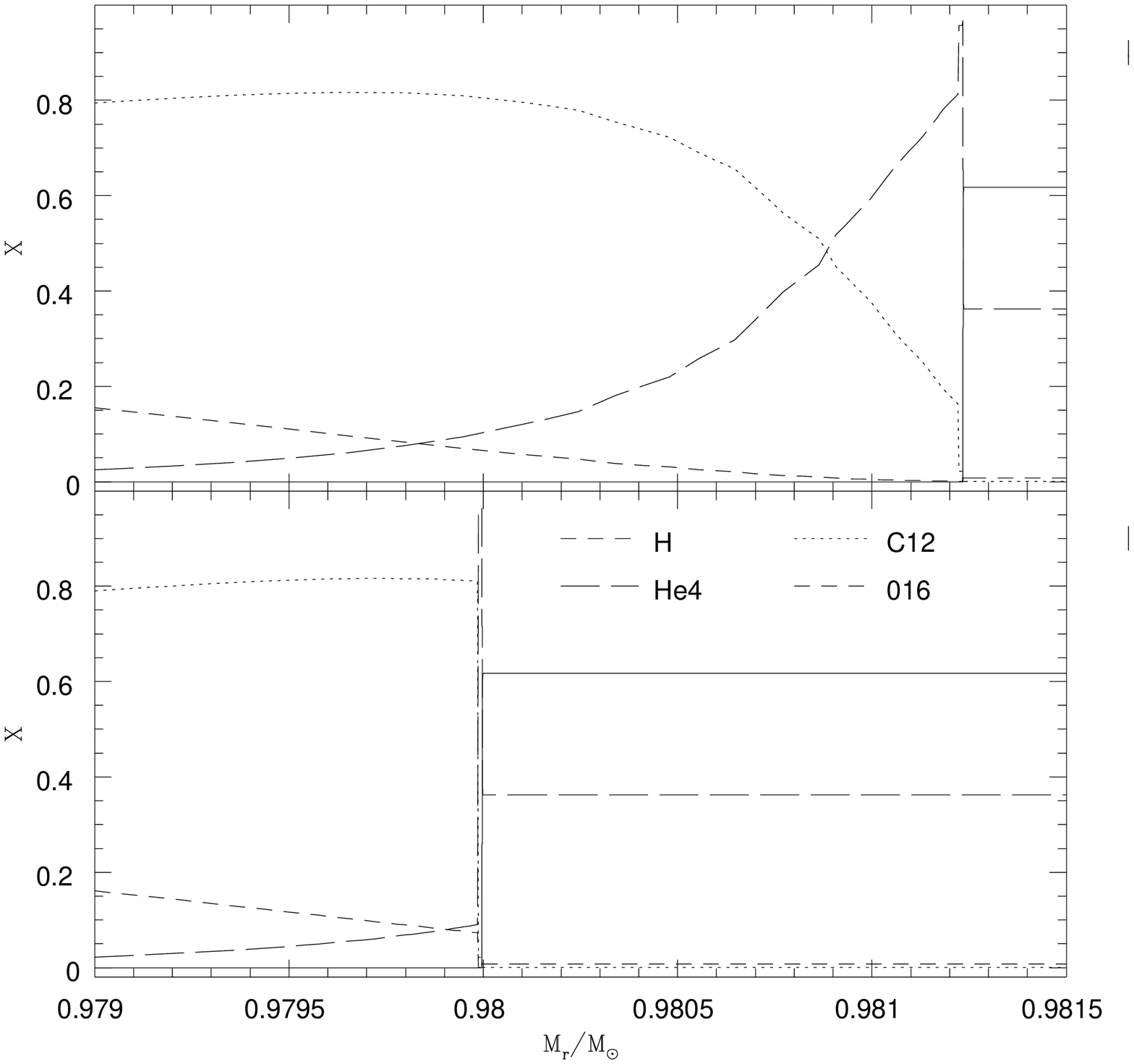,width=8cm}}}
\caption[]{We show the chemistry of two models of 6\Msun\ in late
evolutionary phases, when "symmetric" overshooting is present. The right
figures are blow--up of the left ones close to the H--burning shell.
The dredge--up of carbon and oxygen rich matter from the core to the surface
due to penetration of the external convective envelope through the H--rich
shell is evident.}
\label{fchim}
\end{figure*}

\begin{figure*}
\centerline{\hbox{
\psfig{figure=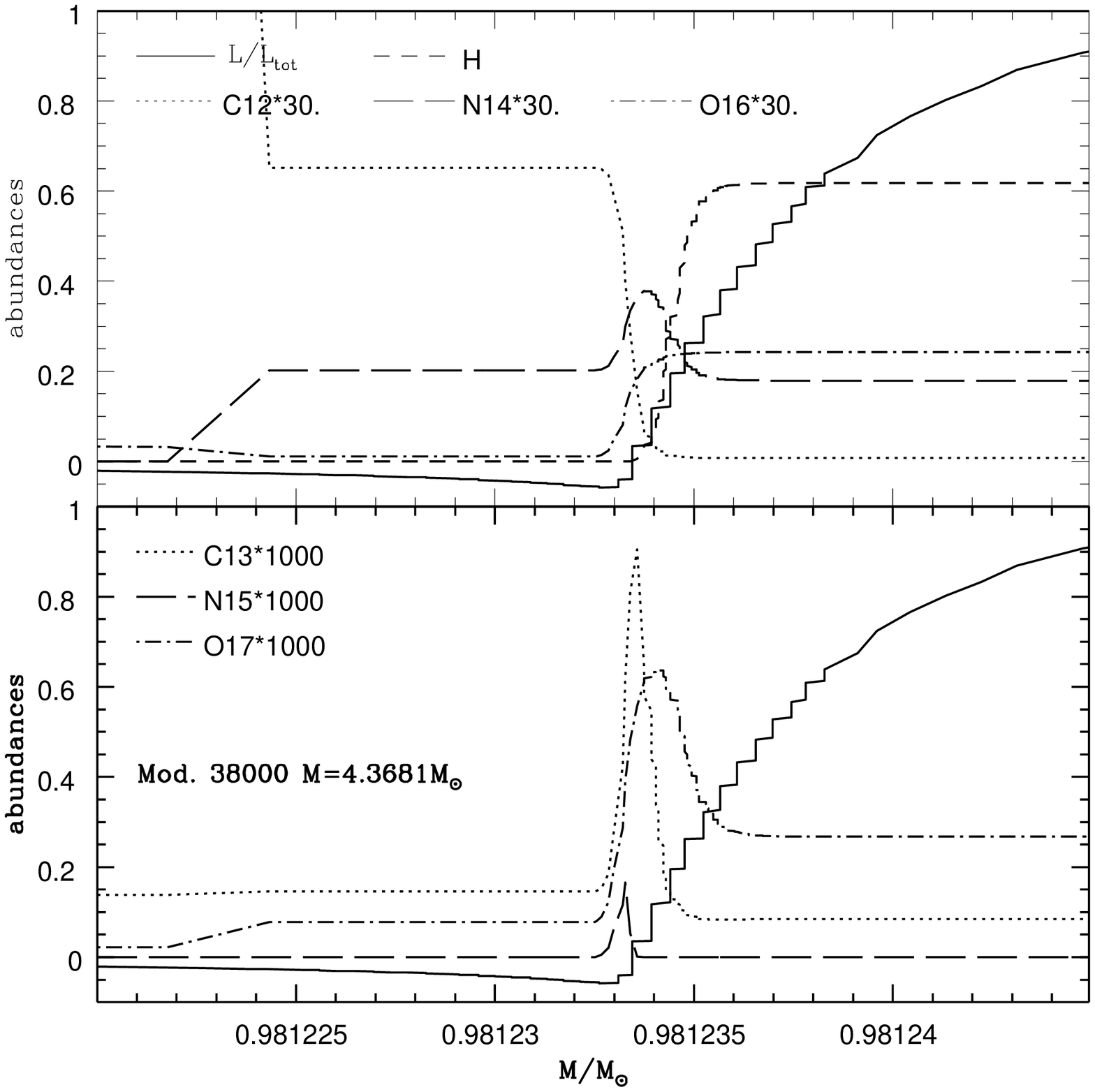,width=8cm}
\psfig{figure=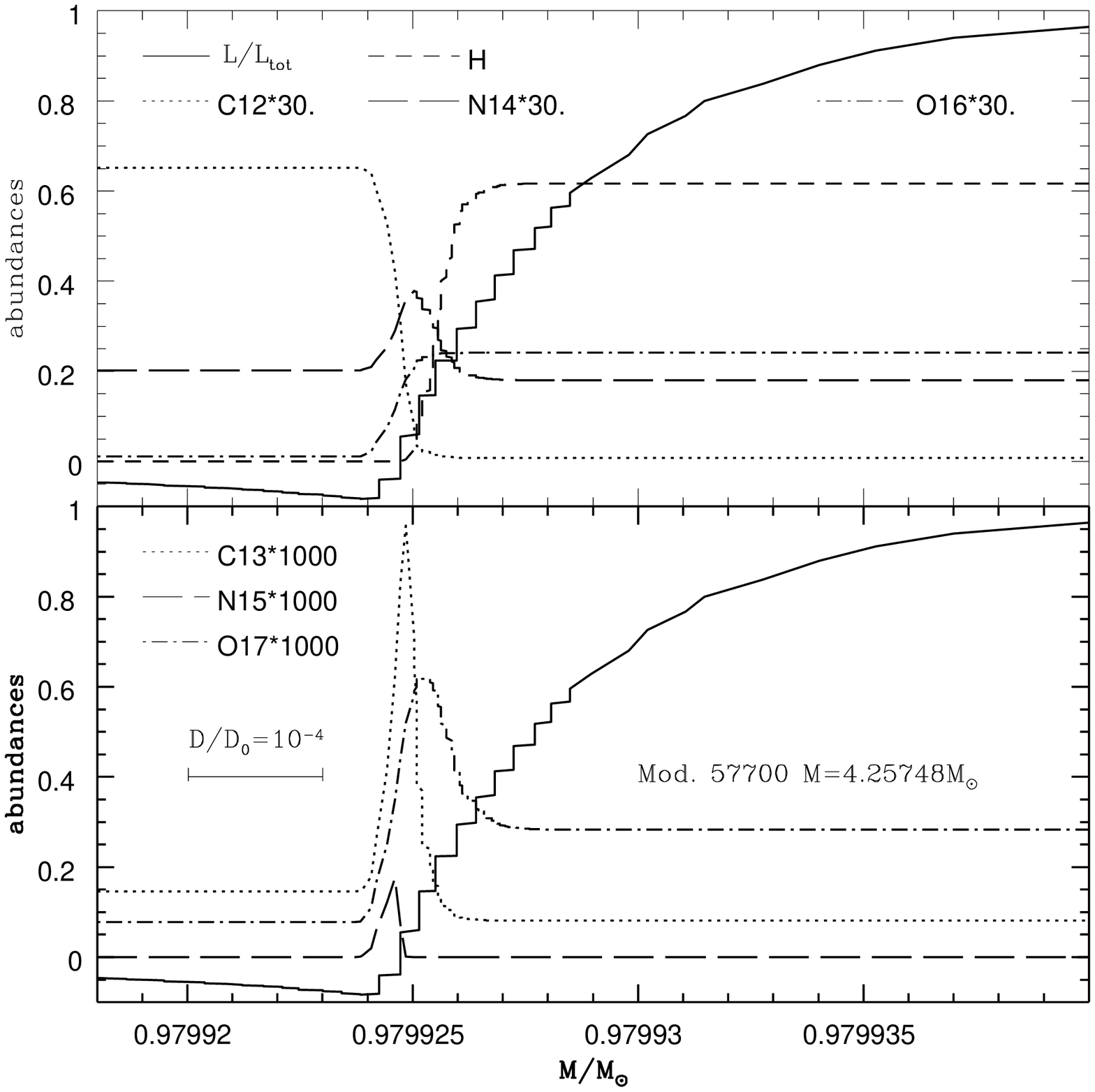,width=8cm}}}
\caption[]{The same as in Fig.13, but further blown up around the
H--burning shell, defined according to the luminosity profile. The hydrogen
profile (dashed) shows that convection is able to homogenize the shell almost
down to its bottom. The CNO elements show however profiles of abundances {\it
inside} the convectively mixed region, which can only be obtained by solving
the coupled diffusive and nuclear evolution. The importance of applying this
latter algorithm could not be better clarified.}
\label{fshell}
\end{figure*}

Notice, in passing, the presence of a $^{13}C$ {\it pocket} just below the
H/C interface, due to the leakage of protons in the C--rich layers. In
principle, $^{13}C$ could act as a neutron source to build up s--process
elements (e.g. Iben 1982), but it must be recalled that TPs are not present
in this case. So, the temperature of the $^{13}C$--rich layers is never
larger than $100 \div 120$ MK, and s--processes operate at a very slow pace.

In summary, according to the above described framework, a possible
observational signature of the presence of overshooting {\it from below} the
convective envelope could arise from:

\begin{itemize}
\item{the practical impossibility of getting a different, straightforward
evolutionary scenario giving rise to low--mass (and low--luminosity) carbon
stars (Herwig et al. 1997);}
\item{the observation that the most massive AGB stars in the Galaxy, selected
as those already having a circumstellar envelope due to mass loss, clearly
show a surface chemistry due to HBB, but no track of s--processed elements
(Garcia Lario et al. 1999);}
\item{the anomalous overabundances of $C$\ and $O$\ in some PG1159 WDs and,
related to this feature:}
\item{the straightforward explanation of the surface exposure of C/O
mixtures in large mass WDs required to explain the fast nova mechanism.}
\end{itemize}

Of course, other explanations are possible for all the above features.
For instance, the lack of s--process enhancements in the Garcia Lario et al.
sample may be due to the fact that pollution of the massive envelope of a
massive AGB stars by s--processes would require a long series of TPs,
while lithium production occurs as soon as the star is on the AGB.
What is relevant is, in our opinion, the fact that several, still {\it
qualitatively} (and not only {\it quantitatively}) unexplained, features,
spontaneoulsly {\bf occur} if {\it symmetric} overshooting is allowed.

\section{Mass dependence}

In the end, we also computed evolutionary sequences for different initial
mass stars, starting from $3.5M_\odot$ and with steps of $0.5M_\odot$. The
chemistry was solar, and both overshooting {\it from above} and mass loss
(respectively: with $\zeta=0.02$ and $\eta=0.1$) were included. Of course,
these computations are still preliminary to a full investigation on lithium
enrichment of the interstellar medium from intermediate mass AGB stars.

The model of $3.5M_{\odot}$ never reaches HBB conditions, and the
Cameron-Fowler mechanism is then not ignited. Also, the same model does not
undergo the second dredge--up, by a small amount. Already at $4.0M_\odot$,
both second dredge--up and HBB are found if overshooting is included, whereas
sequences without overshooting show these features only from $4.5M_\odot$ on.
In all the investigated cases, a strict relation was found between the
presence of second dredge--up, and the following onset of HBB conditions.

Broadly speaking, in all the mass range for which HBB is found, the evolution
follows similar paths at a lower pace the lower is the mass. Also the
amplitudes of the various features connected with HBB (Li--overabundance,
fast increase in luminosity at the beginning of the AGB phase etc.) increase
when increasing the total mass, as one could have expected (Fig.
\ref{fig_14}).
A capillary analysis of the details of the various evolutions is then
useless, and only the overall results of computations of narrower grids of
models, with different masses and chemistries, and with a more realistic
mass loss free parameter ($\eta=0.01\div 0.02$) will be given in a next
paper, as an input for galactic evolution models.

\begin{figure}
\caption[]{For different total masses and models having Z=0.02 and core
overshooting we show the variation with time of the lithium abundance, total
luminosity and
$T_{bce}$.
}
\label{fig_14}
\end{figure}

\begin{figure}
\psfig{figure=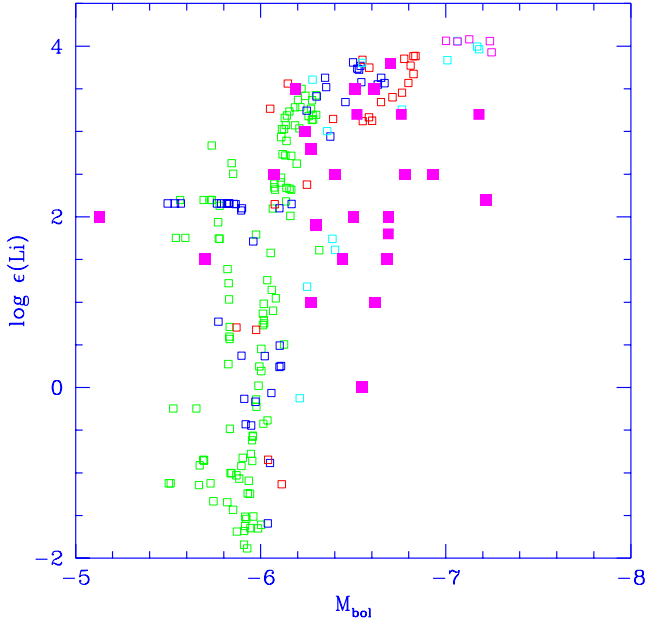,width=8.8cm}
\caption[]{Theoretical positions (open squares) in the lithium vs.
luminosity plane for the 4.0, 4.5, 5.0, 5.5 and 6.0 \Msun models of solar
chemistry with (asymmetric) overshooting and mass loss. Also observed stars
in the magellanic clouds (full squares) are shown.}
\label{caos}
\end{figure}

One last result is instead worth showing. When in AGB, the evolution tends to
become {\it chaotic} in the sense that stars of different masses,
evolutionary phases and pulse/interpulse phases share the same region of the
HR diagram. However, some {\it attractors} can be still identified, as longer
lasting phases. Fig. \ref{caos} shows, in the lithium vs. luminosity plane,
what could be expected from our theoretical evolutionary paths of $4.0 \div
6.0 M_\odot$ stars, with (asymmetric) overshooting and mass loss ($\eta =
0.1$). A mass function $\propto M^{-2}$ has been assumed, and a handful of
points have been spread along the tracks, randomly selected linearly with
time, starting from the beginning of AGB.

Also, a sample of observed stars in the Magellanic Clouds (Smith et al.
1995) are shown in the same figure. Although the metal abundances of these
latter are lower than the theoretical ones, one can identify some first
correspondences. Observed stars of low luminosity are likely to be still
lithium undepleted. The theoretical points mark quite well the left--upper
envelope of the observations. More than that, we have to recall that the
present evolutionary tracks have been followed fon a low number of TPs, and
that models show a trend, when going on with the evolution, to decrease both
the surface lithium abundances and luminosities. In other words, more
complete tracks up to planetary nebula ejection would have led to populate
also the region where the most of the observed stars are present, as also
suggested by the {\it turn--down}s already visible in the upper part of the
theoretical distributions

Of course, no firm conclusions can be still established by these first
results, also because the observational error boxes are quite large, and a
selection effect is certainly present, being easier to detect lithium in the
more luminous objects. However, more extended evolutionary tracks with the
correct chemical imputs (in progress of computation) will probably strengthen
the overall agreement between the theoretical and experimental framework.

\section{Conclusions}

We have examined the details of lithium production in intermediate mass stars
of solar metallicity during the AGB phase. Lithium is naturally produced in
FST models including core overshooting for masses $M\simgt 4M_\odot$, and
$\simgt 4.5 M_\odot$ without overshooting. In MLT models, at least
$6M_{\odot}$ are instead required to ignite HBB with a solar tuning of
$\alpha$, if core overshooting is not allowed. These structures are so
largely dependent on convection, due to the leakadge of the convective
envelope in the H--burning shell, that also other differences (for instance
in the mass -- luminosity relation etc.) between MLT and FST structures are
striking enough that careful comparisons between models and observations
could help constraining the convection model.

The amount of lithium produced during the AGB phase is independent of the
assumed initial lithium abundance, while it is somewhat dependent on the
initial deuterium, and on PMS evolution.

Core overshooting during MS H--burning, leading to larger core masses and
luminosities in AGB, also affects lithium production, which attains a maximum
shortly after the second dredge up. If also overshooting from below the
convective envelope is allowed, in the case of the $6 M_\odot$ star we find
total suppression of the thermal pulses. This mechanism could be responsible
for the presence of high luminosity AGB stars with large lithium
overabundances, but completely lacking the s--elements enhancement associated
to the third dredge--up phase following TPs.

\end{document}